\documentclass[conference]{IEEEtran}
\IEEEoverridecommandlockouts
\usepackage{cite}
\usepackage{amsmath,amssymb,amsfonts}
\usepackage{algorithmic}
\usepackage{graphicx}
\usepackage{textcomp}
\usepackage{subcaption}
\usepackage{url}
\usepackage{xcolor}
\usepackage{multirow}
\usepackage{tabularx}
\usepackage{booktabs}
\def\BibTeX{{\rm B\kern-.05em{\sc i\kern-.025em b}\kern-.08em
    T\kern-.1667em\lower.7ex\hbox{E}\kern-.125emX}}
\begin{document}

\title{MarketSenseAI 2.0: Enhancing Stock Analysis through LLM Agents\\
\thanks{Part of this research was funded by European Union Commission through Project FAME with grant number 101092639.}
}


\author{\IEEEauthorblockN{George Fatouros\textsuperscript{1,2}, Kostas Metaxas\textsuperscript{1,3}, John Soldatos\textsuperscript{2}, Manos Karathanassis\textsuperscript{3}}
\IEEEauthorblockA{\textsuperscript{1}Alpha Tensor Technologies\\
\textsuperscript{2}Innov-Acts\\
\textsuperscript{3}KM Cube Asset Management\\
\{george, kostas\}@alpha-tensor.ai, \{gfatouros, jsoldat\}@innov-acts.com, \{kostas, manos\}@km3am.com}
}


\maketitle

\begin{abstract}
MarketSenseAI is a novel framework for holistic stock analysis which leverages Large Language Models (LLMs) to process financial news, historical prices, company fundamentals and the macroeconomic environment to support decision making in stock analysis and selection. In this paper, we present the latest advancements on MarketSenseAI, driven by rapid technological expansion in LLMs. Through a novel architecture combining Retrieval-Augmented Generation and LLM agents, the framework processes SEC filings and earnings calls, while enriching macroeconomic analysis through systematic processing of diverse institutional reports. We demonstrate a significant improvement in fundamental analysis accuracy over the previous version. Empirical evaluation on S\&P 100 stocks over two years (2023-2024) shows MarketSenseAI achieving cumulative returns of 125.9\% compared to the index return of 73.5\%, while maintaining comparable risk profiles. Further validation on S\&P 500 stocks during 2024 demonstrates the framework's scalability, delivering a 33.8\% higher Sortino ratio than the market. This work marks a significant advancement in applying LLM technology to financial analysis, offering insights into the robustness of LLM-driven investment strategies.
\end{abstract}


\section{Introduction}
MarketSenseAI\footnote{MarketSenseAI is available at \url{https://www.marketsense-ai.com/}} is a holistic framework designed to leverage Large Language Models (LLMs) for stock analysis and selection. By processing financial news, historical stock prices, company fundamentals, and macroeconomic data, it aims to support multifaceted decision-making processes in modern financial markets. Since its initial inception \cite{fatouros2024can}, the framework has evolved in tandem with rapid advancements in LLM technology, introducing enhanced capabilities for data-driven investment strategies.

The motivation for developing MarketSenseAI arises from the limitations of existing systematic stock analysis approaches. Many methods rely on time-series modeling, sometimes supplemented by sentiment indicators, yet seldom integrate the broad scope of available data. A significant challenge also lies in handling data with varying sampling frequencies: macroeconomic indicators and fundamental factors typically follow lower-frequency release schedules than market data, requiring sophisticated integration methods to ensure consistency.

Although AI-based solutions employing machine learning or deep learning provide systematic frameworks for stock prediction \cite{bacco2024investigating,sonkavde2023forecasting,fatouros2023deepvar}, they often focus on isolated data types (e.g., sentiment or historical returns) without leveraging the wealth of relevant financial texts as well as the context of those texts. Consequently, investment strategies frequently emphasize price trends, fundamental ratios, or macroeconomic variables in isolation, overlooking the collective dependencies among these factors \cite{aqr2023,MacrosynergyFundamentalValue}. Unlike traditional quantitative models that operate as black boxes, MarketSenseAI supplies detailed explanations for its investment decisions, thereby enhancing transparency and user trust \cite{mavrepis2024xai}.

Even approaches that incorporate textual data (e.g., news or earnings call transcripts) tend to center on predicting sentiment indicators rather than conducting in-depth qualitative analysis \cite{fatouros2023transforming,nti2020systematic}. This fragmentation is further compounded by limited human resources for processing heterogeneous financial information at scale. In this context, integrating structured financial data with unstructured financial information becomes a challenge—one that MarketSenseAI seeks to address.

However, the successful application of LLMs in finance also poses notable challenges. First, even state of the art LLMs have constraints on context window size, limiting their ability to process large documents such as 10-K filings or detailed macroeconomic reports \cite{wang2024prompt, li2024long}. Second, model outputs can be sensitive to prompt engineering choices and broader design decisions, complicating issues such as backtesting and replicability \cite{yang2024harnessing}. Third, consistently interpreting—and accurately handling—quantitative metrics like risk measures and financial ratios can be difficult due to the probabilistic nature of LLMs \cite{zheng2024large, mirzadeh2024gsm}. Additionally, ensuring models remain current with newly released data is non-trivial, particularly as most pre-trained LLMs have fixed cut-off dates \cite{liu2024llms}.

In response to these challenges, the contributions of this paper focus on demonstrating how recent advances in LLM architectures can strengthen fundamental and macroeconomic analyses within the MarketSenseAI framework:

\begin{enumerate} 
    \item \textbf{Refined Fundamental Analysis}: We introduce a Chain-of-Agents (CoA) approach that enables granular handling of large-scale financial data—such as 10-Q, 10-K reports, and earnings call transcripts—to deliver more accurate assessments of a company’s financial standing. 
    \item \textbf{Enhanced Macroeconomic Analysis}: A dedicated Retrieval-Augmented Generation (RAG) module, employing semantic chunking and Hypothetical Dense Embeddings (HyDE)-based retrieval, processes a broader range of expert reports and indicators, providing the macroeconomic context often missing in traditional analytics. 
    \item \textbf{Detailed Real-World Evaluation}: Experiments using S\&P 100 stocks for a two-year period (2023–2024) and S\&P 500 stocks for 2024 illustrate the robustness of our proposed system, revealing a notable improvement in fundamental analysis accuracy and consistent excess returns of 8.0–18.9\% with comparable risk over benchmark indices.
\end{enumerate}

These enhancements position MarketSenseAI as a candidate for both retail and institutional investors seeking advanced analytics. By merging multiple data streams and applying specialized LLM agents, MarketSenseAI demonstrates how AI-driven strategies can yield improved investment recommendations and deeper market insights.

The remainder of this paper is structured as follows: Section~\ref{sec:2} provides a literature review examining current research in LLM-based systems for financial analysis. Section~\ref{sec:3} details updates to the MarketSenseAI architecture, including agent responsibilities and data flow. Section~\ref{sec:4} presents our experimental design, covering datasets, evaluation metrics, and baseline comparisons. Section~\ref{sec:5} discusses empirical findings from S\&P 100 and S\&P 500 stocks, including performance metrics, risk-adjusted returns, and a factor analysis. Finally, Section~\ref{sec:6} concludes with key insights and outlines future developments for MarketSenseAI.

\section{Background and Related Work}
\label{sec:2}

Recent advances in LLMs have spurred a wave of research into their applicability to diverse financial tasks, including fundamental analysis, alpha discovery, and portfolio decision-making. This section surveys closely related work in four main areas: (i) LLM-based fundamental analysis, (ii) advanced methods in LLM-driven investment analysis, (iii) retrieval-augmented techniques, (iv) the significance of SEC filings and earnings conference calls in fundamental research, and (v) the impact of the macroeconomic environment on stock analysis.

\subsection{LLM-Based Fundamental Analysis}
A growing body of literature investigates how LLMs can replicate or surpass human analysts’ capabilities for parsing and interpreting financial statements. For instance, \cite{kim2024financial} demonstrate that GPT-4 can execute ratio analysis and detect trends via Chain-of-Thought (CoT) prompting \cite{wei2022chain}, yielding interpretable explanations and confidence assessments for binary earnings forecasts. Similarly, \cite{cheng2024gpt} employ GPT-4 to generate high-return factors grounded in economic reasoning, thereby laying a foundation for quantitative investment models. Both studies highlight LLMs’ ability to extract structured insights, such as financial ratios and performance patterns, directly from extensive textual documents.

\subsection{Advanced Methods in LLM-Driven Investment Analysis}
Beyond processing financial disclosures, LLMs have also been employed to generate alpha signals and optimize trading strategies. \cite{wang2023alpha} introduce Alpha-GPT, which couples human expertise with automated alpha discovery to refine trading signals. Similarly, TradingGPT \cite{li2023tradinggpt} adopts a multi-agent, layered memory architecture for collaborative decision-making—though its evaluation results are limited. Meanwhile, \cite{tan2023large} apply sentiment analysis, model ensembles, and in-context learning to predict returns in the Chinese equity market, achieving promising accuracy. More recently, \cite{papasotiriou2024ai} demonstrate that GPT-4, leveraged through in-context learning, can produce stock ratings (e.g., buy, hold, sell) from fundamental reports and news data—outperforming human analysts in certain scenarios.

\subsection{Retrieval-Augmented Techniques}
RAG \cite{lewis2020retrieval} has emerged as one of the most prevalent applications of LLMs in production systems \cite{menlo_ventures_2024}, allowing models to incorporate extensive corpora beyond their internal parameters and input context. This approach is particularly valuable for finance, where multi-faceted data—regulatory filings, market news, economic reports—can be vast and continually updated. Recent research focuses on advanced chunking, query expansion, and re-ranking algorithms to mitigate context loss when processing large documents \cite{setty2024improving, yepes2024financial}, though optimal methodologies may vary depending on data size, structure, and recency requirements. For instance, in stock analysis, the date-aware document retrieval becomes essential yet is often overlooked in standard similarity searches. Although a few recent works propose RAG pipelines tailored to financial tasks \cite{arslan2024business, zhang2023enhancing}, there remains a gap in comprehensive, domain-specific solutions optimized for financial analytics.

\subsection{Importance of Filings and Earnings Calls in Fundamental Research}
A substantial body of empirical evidence underscores the critical role of SEC filings (e.g., 10-K and 10-Q) and earnings conference calls in shaping market outcomes and guiding investment decisions. Studies by \cite{loughran2011liability, eugene1992relationship} report that changes in language complexity, disclosure content, and tonal shifts within filings predict returns, risk profiles, and management quality. \cite{campbell2008search, mayew2015md} emphasize the importance of footnote analysis for identifying hidden risks, while \cite{dikolli2019cfo} demonstrate how readability and clarity can serve as proxies for managerial competence and earnings transparency.

Earnings conference calls exert a similarly influential role in price discovery. \cite{frankel1999empirical} find that trading volumes and volatility spike during these events, especially in Q\&A sessions where spontaneous managerial insights can move markets. \cite{mayew2012power} show that  the tone of calls offer predictive power regarding a firm’s future performance, while \cite{price2012earnings} reveal how the qualitative tone of calls influences both subsequent returns and analyst revisions. \cite{li2008annual, larcker2012detecting} note that these qualitative cues provide additional signals beyond quantitative metrics, and may even reveal deceptive statements. Finally, \cite{mayew2015md} document how analysts with direct access to earnings calls can generate more precise forecasts. Together, these findings establish filings and conference calls as indispensable avenues for uncovering deeper insights into a firm’s performance and strategy.

Emerging research highlights transformative potential of LLMs in financial disclosures and analysis. For instance tools like ChatReport \cite{ni2023chatreport} and XBRL-Agent \cite{han2024xbrl} show LLMs can democratize analysis of dense reports through automated extraction of sustainability metrics and financial concepts, though challenges persist in numerical accuracy and hallucination mitigation. \cite{cook2023evaluating} validate LLMs’ viability in parsing earnings call sentiment, while \cite{goldsack2024facts} reveal their capacity to generate multi-perspective analytical reports approaching human quality. These advances suggest LLMs could reshape fundamental analysis workflows, but require careful governance to preserve informational integrity.

\subsection{Macroeconomic environment impact in stock analysis}
While fundamental metrics and firm-specific disclosures remain critical, macroeconomic indicators (e.g., GDP growth, inflation rates, interest rates), central bank policies, geopolitical factors, and trade agreements between nations provide a broader context that can significantly influence investment outcomes \cite{kwon2024large}. Fluctuations in these external conditions can affect corporate earnings, valuation models, and overall market sentiment—ultimately impacting both short- and long-term trading strategies.

Expert analysis from leading financial institutions plays a crucial role in interpreting these complex macroeconomic relationships. Research and opinionated reports from investment banks and central banks provide valuable insights into emerging trends, policy implications, and potential market impacts that may not be immediately apparent in quantitative data alone \cite{abaidoo2023inflation}. These expert opinions are particularly valuable when analyzing interconnected global markets where local expertise and institutional knowledge become essential for understanding market dynamics.

Incorporating macro-level context and expert insights alongside firm-level data can lead to more robust and adaptive models, particularly when combined with LLM-based frameworks capable of integrating multiple data streams. Notably, macroeconomic forces often vary in their impact across different stocks and sectors. For example, US tariffs on imported goods from China can weigh heavily on industries reliant on specific commodities or products \cite{nvidia2022q3}. However, many existing quantitative and LLM-based stock-analysis models typically overlook these broader economic factors and expert interpretations, revealing a gap in current approaches to investment research.

\section{Methods}
\label{sec:3}

\subsection{Overview of MarketSenseAI components}
\label{sec:3.1}

\begin{figure}[htbp]
\centerline{\includegraphics[width=0.5\textwidth]{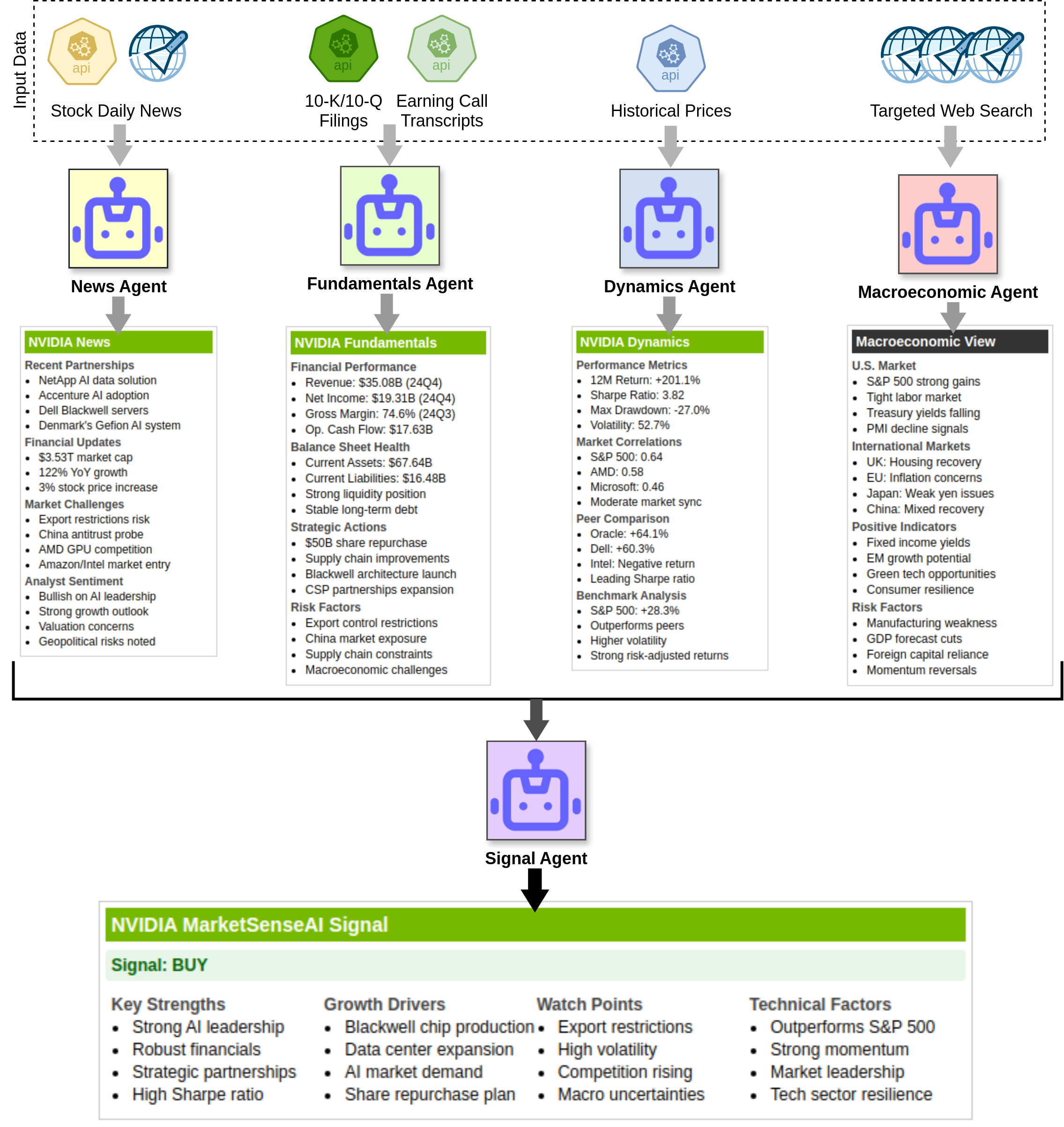}}
\caption{Conceptual Architecture of MarketSenseAI, highlighting for a selected stock (i.e., Nvidia on Jan. 3, 2025). The agents' outputs have been condensed for illustration purposes.}
\label{fig:marketsense}
\end{figure}

The MarketSenseAI framework, detailed in \cite{fatouros2024can}, is designed as a modular system that synthesizes various types of financial information—from daily news and corporate fundamentals to market dynamics and macroeconomic data—to generate actionable investment signals. As shown in Fig.~\ref{fig:marketsense} the system consists of five primary LLM agents:

\begin{enumerate}
    \item \textbf{News Agent}: Responsible for aggregating and condensing relevant news articles pertaining to a given stock. Each day’s raw text is first distilled into a concise summary, which is then integrated with previous summaries to form a progressive narrative of recent developments. This mechanism ensures that older but still-relevant news (e.g., open legal cases) remains part of the evolving context.
    \item \textbf{Fundamentals Agent}: Focuses on analyzing each company’s financial statements (e.g., balance sheets, income statements, and cash flow reports). To handle large and often complex numerical data, these statements are preprocessed and reduced into abbreviated formats (e.g., grouping figures in “million” or “billion”) before the LLM extracts key insights. The system compares recent quarters to highlight shifts in profitability, revenue, or leverage ratios, laying the groundwork for fundamental analysis. Besides the numerical figures, the updated agent analyzes SEC filings and earnings call transcripts as analyzed at Section~\ref{sec:3.2}.
    \item \textbf{Dynamics Agent}: Examines historical price movements and contextualizes them against industry peers and the broader market (i.e., S\&P~500). By incorporating risk metrics like volatility, Sharpe Ratio, and maximum drawdown statistics, this component provides a risk-adjusted lens on how the target stock performs relative to both its closest competitors and the general market.
    \item \textbf{Macroeconomic Agent}: Collates and synthesizes key macro-level reports, including investment bank outlooks, central bank announcements, and broader geopolitical or sector-specific research. The generated summary distills multiple sources into a concise snapshot of prevailing economic conditions (e.g., interest rate policies, inflation trends, and global demand shifts). The resulting macro-level insight helps the system account for external forces that may affect individual stocks or entire sectors.
    \item \textbf{Signal Agent}: The final component integrates the textual outputs from the previous four modules—news, fundamentals, price dynamics, and macroeconomic outlook—into a single decision-making process. Implemented via a CoT prompting strategy, the LLM reviews each aggregated summary to produce an investment signal (\textit{buy}, \textit{hold}, or \textit{sell}). It also provides a written explanation that traces the reasoning behind each recommendation, thereby enhancing transparency and interpretability.
\end{enumerate}

Each of these components can be run and leveraged by stakeholders independently. This modularity not only allows new information sources to be plugged in but also enables flexibility in how data are refreshed (e.g., daily news versus quarterly fundamentals). 

\subsection{Enhanced fundamentals analysis}
\label{sec:3.2}

\begin{figure}
    \includegraphics[width=0.5\textwidth]{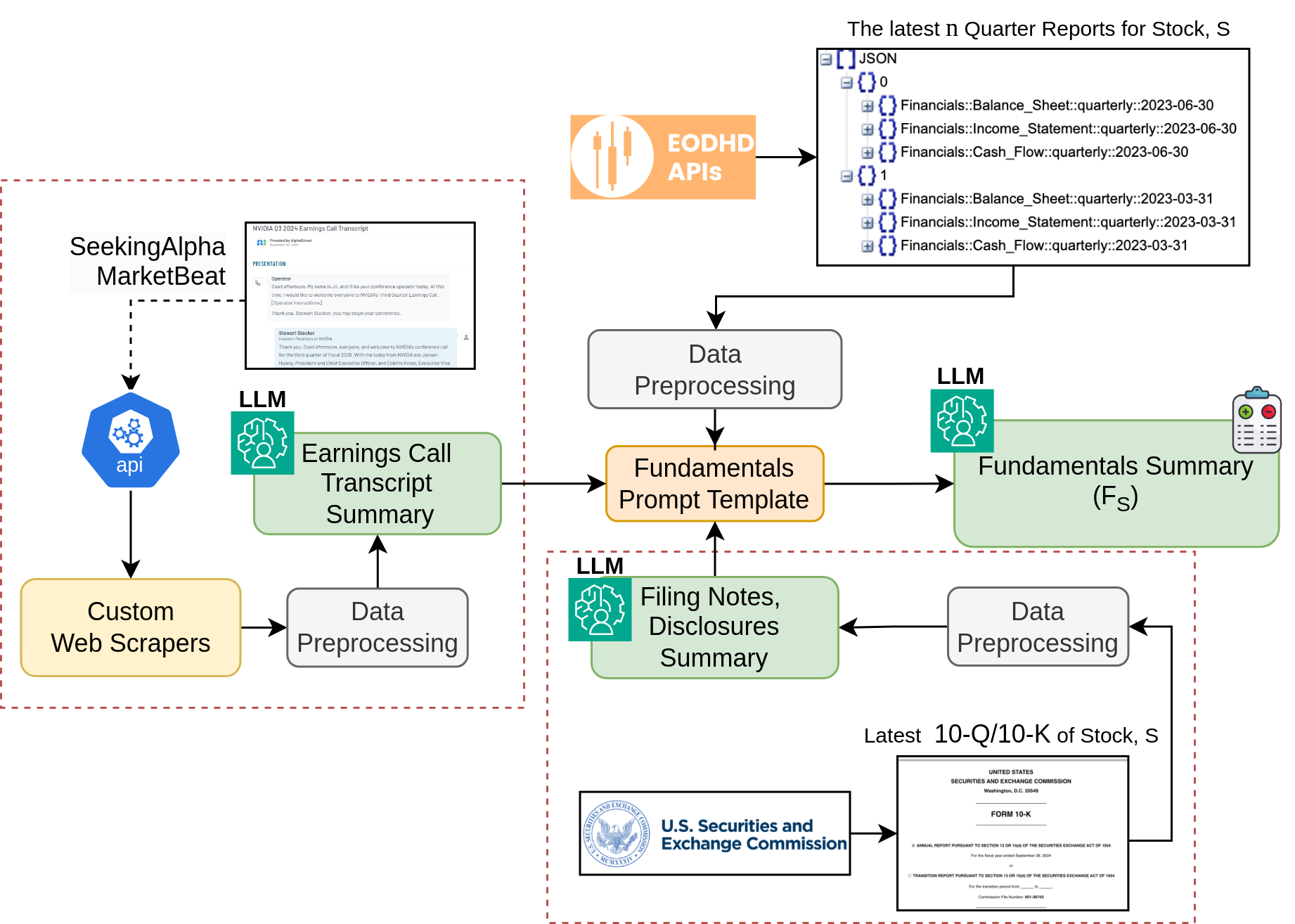}
    \caption{Fundamentals Agent architecture. Processes in red boxes depict the new processes responsible for integrating the company notes and disclosures from SEC filings and insights from earning call's press conference.} 
    \label{fig:fundamentals}
\end{figure}

The Fundamentals Agent in MarketSenseAI has been significantly enhanced to go beyond the numerical analysis of financial statements by incorporating three sequential LLM processes (Fig.~\ref{fig:fundamentals}). While the previous version focused primarily on extracting trends and ratios from standard reports (e.g., balance sheets, income statements, and cash flow statements), the updated agent now processes disclosures, footnotes, and strategic insights found in 10-Q and 10-K SEC filings. Moreover, it accounts for the qualitative dimension of earnings call transcripts, including their Q\&A sessions. These additions enable deeper context and transparency by capturing forward-looking guidance, managerial tone, and strategic outlooks that are not apparent from numerical data alone.

\subsubsection{A Three-Layer Approach to Integrating Qualitative and Quantitative Data} 
\label{sec:3.2.1} 

To generate a holistic fundamental summary for a given company, the agent orchestrates three primary LLM processes:

\begin{enumerate} 

\item \textbf{Filing Summary}: Textual information from SEC filings is summarized with particular emphasis on disclosures, risk factors, and strategic initiatives. These elements help explain the reasons behind fluctuations or significant changes in key financial metrics.
\item \textbf{Earnings Call Summary}: Earnings call transcripts are processed separately to extract management’s qualitative signals, such as sentiment, confidence, and forward-looking statements. This phase focuses on the executive team’s tone, discussions on partnerships or product launches, and any macro-level considerations that may influence long-term performance.
\item \textbf{Fundamental Consolidation}: The outputs from the first two processes are combined with the latest five quarters of numerical data—covering profitability, revenue growth, debt levels, cash flow, and liquidity—into a final LLM task. This consolidated analysis delivers a cohesive narrative, one that not only summarizes the quantitative metrics but also contextualizes them with the insights gleaned from the filings and earnings call.
\end{enumerate}

Compared to the previous version of MarketSenseAI, this multi-stage method ensures that both factual and interpretive aspects of a company’s financial health are captured. The agent can now highlight the drivers behind profit surges or downturns, discuss newly disclosed risks, and evaluate potential shifts in management strategy.

\subsubsection{Evaluating the Impact of SEC Filings and Earnings Calls} \label{sec:3.2.2} 

\begin{figure}[!t]
\centering
\begin{subfigure}{0.48\columnwidth}
\centering
\includegraphics[width=\textwidth]{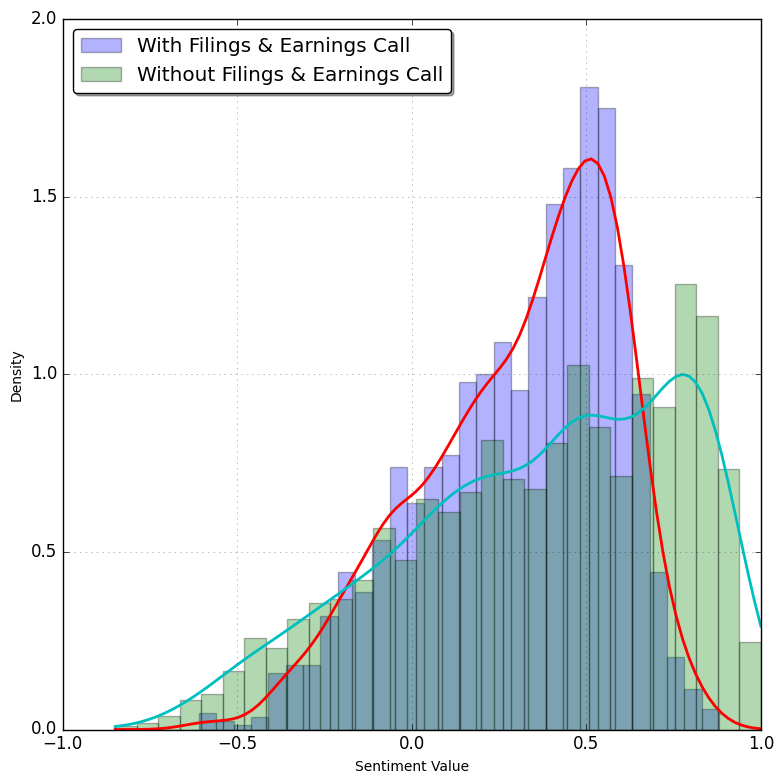}
\caption{}
\label{fig:fundamentals_hist}
\end{subfigure}
\hfil
\begin{subfigure}{0.48\columnwidth}
\centering
\includegraphics[width=\textwidth]{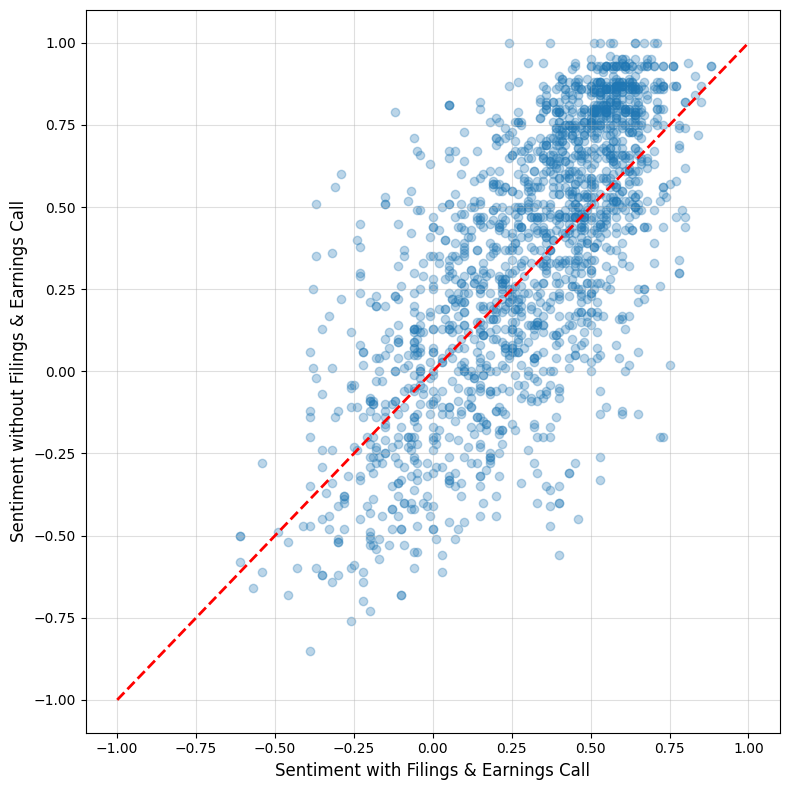}
\caption{}
\label{fig:fundamentals_scatter}
\end{subfigure}
\caption{Analysis of Fundamentals Agent's sentiment output: (a) histogram distribution and (b) scatter plot comparison. Points below the line indicate cases where sentiment improved after incorporating filings and earnings call data.}
\label{fig:fundamentals_analysis}
\end{figure}

To assess how SEC filings and earnings call data affect the Fundamentals Agent's outputs, we conducted a sentiment analysis on 1,500 generated summaries covering S\&P500 stocks at three different points in time. The FinBERT model was utilized to obtain the sentiment of each generated summary \cite{huang2023finbert}.  Table\ref{tab:sentiment_stats} and Fig.~\ref{fig:fundamentals_analysis} reveal distinct patterns between outputs with and without this additional text-based information. When incorporating filings and calls data, the analysis showed a slightly less positive average sentiment (Mean = 0.31) with more moderate variance (Std Dev = 0.28). In contrast, analyses based on numerical data alone exhibited more positively skewed results (Mean = 0.36) with a wider spread of sentiment values (Std Dev = 0.40). This moderation in sentiment when including filings data is particularly noteworthy as SEC filings require companies to disclosure risks and uncertainties in dedicated sections, even when their financial metrics appear strong, thus providing a more balanced perspective of the company's outlook.

Although the two setups differ in their sentiment distributions, the variability in scores underscores how qualitative insights can moderate an otherwise upbeat narrative based solely on numerical trends. Notably, the mean difference of 0.24 and a maximum difference of 0.96 suggest that incorporating the text from filings and calls can reveal otherwise unrecognized risks or strategic realignments.

\begin{table}[htbp]
\caption{Statistics of sentiment analysis of Fundamentals Agent's output across stocks and dates.}
\begin{center}
\tiny
\begin{tabular}{|l|c|c|c|}
\hline
\textbf{Statistic} & \textbf{Sentiment (Full)$^{\mathrm{a}}$} & \textbf{Sentiment (Basic)$^{\mathrm{b}}$} & \textbf{Difference} \\
\hline
Mean & 0.31 & 0.36 & 0.24 \\
\hline
Std. Dev. & 0.28 & 0.40 & 0.17 \\
\hline
Minimum & -0.61 & -0.85 & 0.00 \\
\hline
25$^{\text{th}}$ Percentile & 0.12 & 0.08 & 0.10 \\
\hline
Median & 0.37 & 0.44 & 0.21 \\
\hline
75$^{\text{th}}$ Percentile & 0.53 & 0.72 & 0.33 \\
\hline
Maximum & 0.88 & 1.00 & 0.96 \\
\hline
\multicolumn{4}{l}{$^{\mathrm{a}}$Full includes SEC Filings and Earnings Call transcripts.} \\
\multicolumn{4}{l}{$^{\mathrm{b}}$Basic includes only numerical data from quarterly statements.}
\end{tabular}
\label{tab:sentiment_stats}
\end{center}
\end{table}

\begin{figure}[!t]
\centering
\begin{subfigure}{0.48\columnwidth}
\centering
\includegraphics[width=\textwidth]{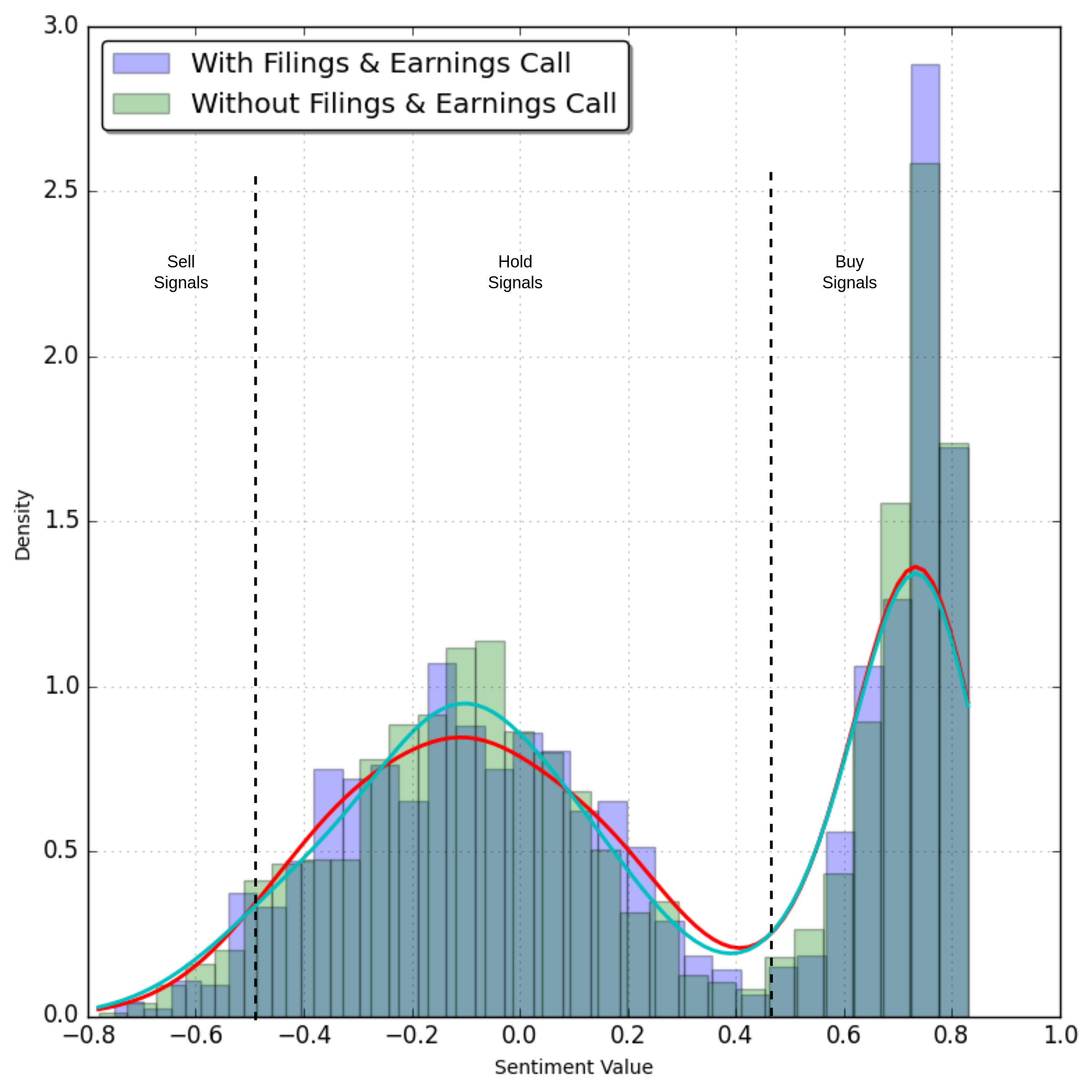}
\caption{}
\label{fig:ms_hist}
\end{subfigure}
\hfil
\begin{subfigure}{0.48\columnwidth}
\centering
\includegraphics[width=\textwidth]{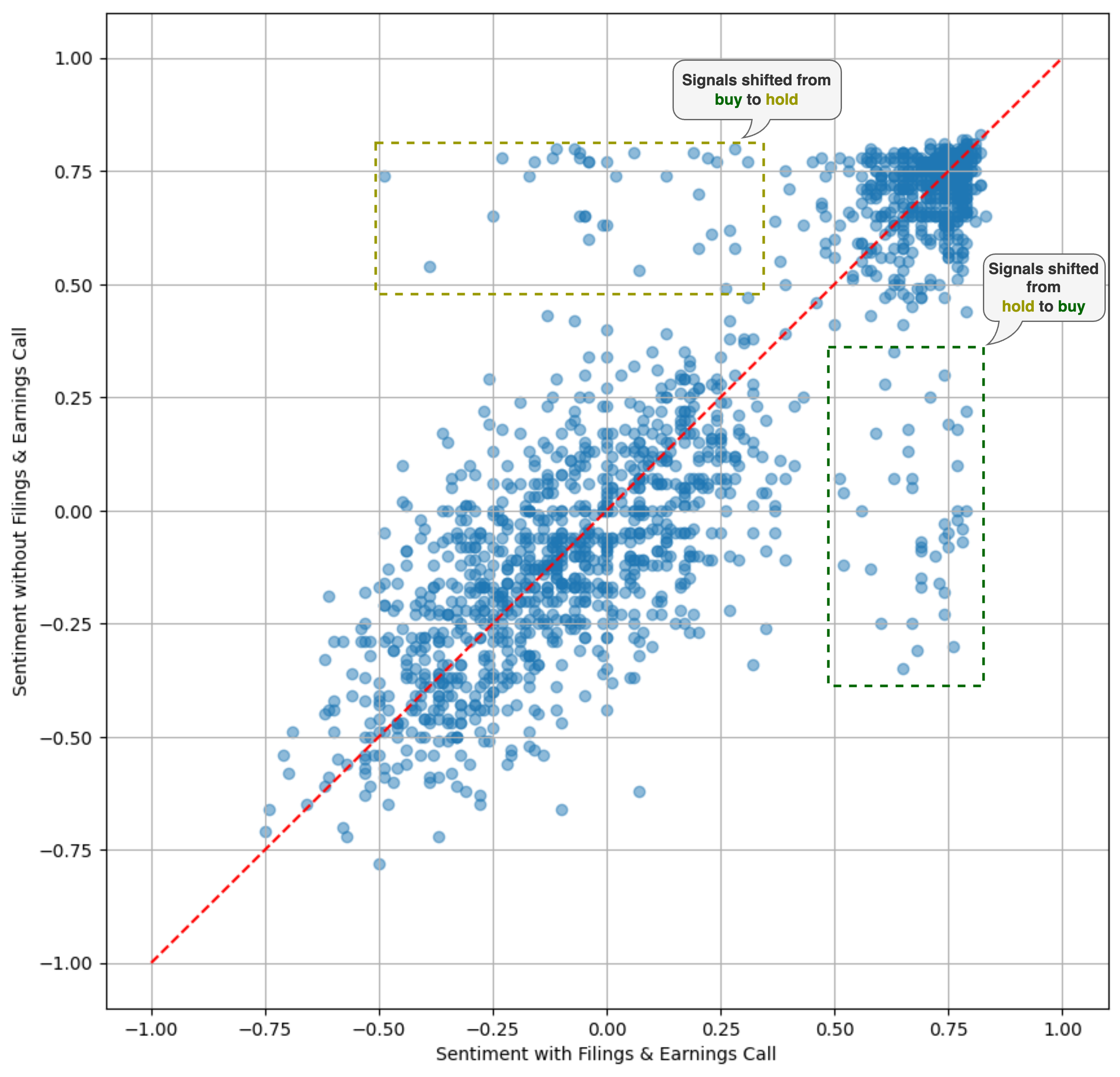}
\caption{}
\label{fig:ms_scatter}
\end{subfigure}
\caption{Analysis of Signal Agent's sentiment output: (a) histogram distribution and (b) scatter plot comparison. Points in the yellow and green boxes indicate cases where the incorporation of filings and earnings call data results in a change of the stock signal.}
\label{fig:ms_analysis}
\end{figure}

We also investigated how this enhanced Fundamentals Agent influences \emph{final} investment signals in MarketSenseAI (Fig.~\ref{fig:ms_analysis}). While the overall distribution of text sentiment in the system’s signal explanations remains consistent (\ref{fig:ms_hist}), roughly \emph{5\%} of signals were downgraded from \emph{buy} to \emph{hold} or upgraded from \emph{sell} to \emph{hold} once the system considered insights from the filings and earning calls  (\ref{fig:ms_scatter}). This outcome shows that combining qualitative context with quantitative metrics produces a more complete assessment, one that can shift investment recommendations in the presence of textual information.

Taken together, these results highlight the updated Fundamentals Agent’s ability to integrate domain-specific textual sources to generate more insightful analyses. By incorporating details on forward-looking statements, strategy, and potential pitfalls, the agent ensures that generated recommendations are grounded in a broader, more comprehensive understanding of each company’s position and prospects.

\subsection{Macroeconomic Analysis Improvements}
\label{sec:3.3}

The Macroeconomic Agent, which functions as an economist within MarketSenseAI, has been enhanced to process a broader range of institutional reports through a robust data-ingestion and generation pipeline (Fig.~\ref{fig:macro1} and \ref{fig:macro2}). These updates address known limitations of LLMs such as constrained context windows, the tendency to generate hallucinations, and oversimplification—by systematically incorporating diverse macroeconomic data from authoritative sources. As a result, the Macroeconomic Agent can provide more comprehensive and context-rich analysis on factors that influence stock performance.

\subsubsection{Data Injection}
\label{sec:3.3.1}

\begin{figure}[htbp]
\centerline{\includegraphics[width=0.5\textwidth]{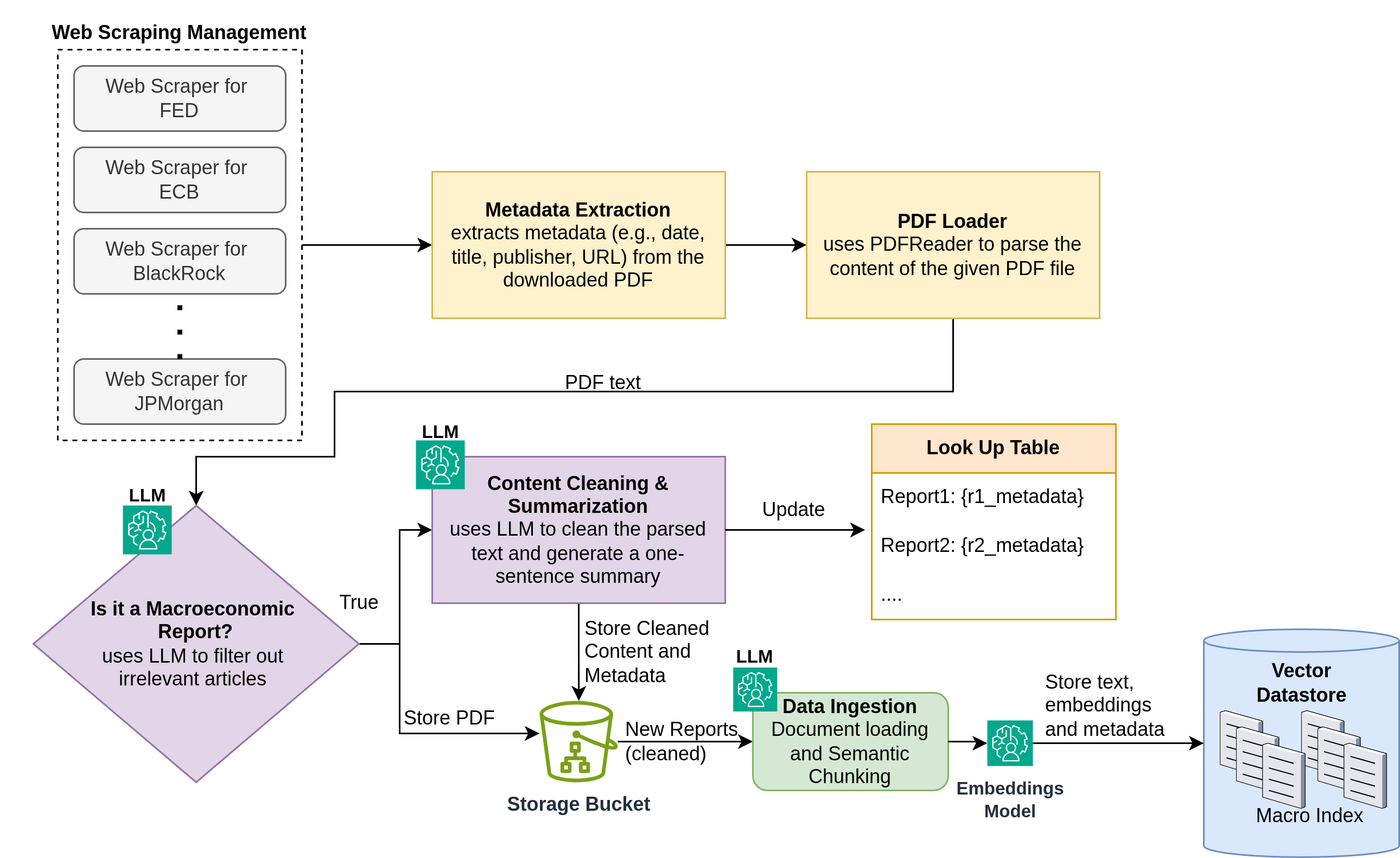}}
    \caption{Macroeconomic Agent's functions during data injection.} 
    \label{fig:macro1}
\end{figure}

The data injection stage (Fig.~\ref{fig:macro1}) is designed to efficiently collect, process, and store macroeconomic reports from multiple sources, including central banks (e.g., FED, ECB), statistical bureaus, the International Monetary Fund, the Bank for International Settlements, and sell-side reports from global investment banks such as JPMorgan and BlackRock. We have implemented institution-specific parsing scripts that handle the unique formatting and structure of reports from each source, ensuring consistent and accurate data extraction across different providers.

\textbf{Metadata Extraction and Filtering:} Once a document is parsed, we identify key attributes like publication date, publisher, and URL. These metadata not only ensure document provenance but also enables the system to sequence reports chronologically. Next, an LLM-powered classifier determines whether the text is relevant to macroeconomic analysis. Irrelevant documents (e.g., marketing brochures) are discarded at this step.

\textbf{Content Cleaning and Summarization:} For relevant documents, another LLM process removes extraneous text (e.g., disclaimers, duplicate headers) and produces a summary capturing the document’s core insights. Large files (over 30 pages) are broken into smaller chunks; each chunk is cleaned, summarized, and then consolidated into a single refined representation of the entire document. This approach preserves vital macroeconomic details without overwhelming LLM context limits.

\textbf{Storage and Indexing:} The cleaned content, along with metadata, is stored. Parallelly, a look up table is updated with relevant metadata to maintain an organized inventory of all processed documents. Afterward, we conduct semantic chunking of new reports \cite{zhao2024meta}; each chunk is embedded and stored in a Vector Datastore for fast, similarity-based retrieval. By chunking on natural boundaries (e.g., the end of a section or a shift in economic theme), the system ensures granular and semantically coherent indexing of macroeconomic information \cite{taipalus2024vector}.

\subsubsection{Macroeconomic Data Generation}
\label{sec:3.3.2}

\begin{figure}[htbp]
\centerline{\includegraphics[width=0.5\textwidth]{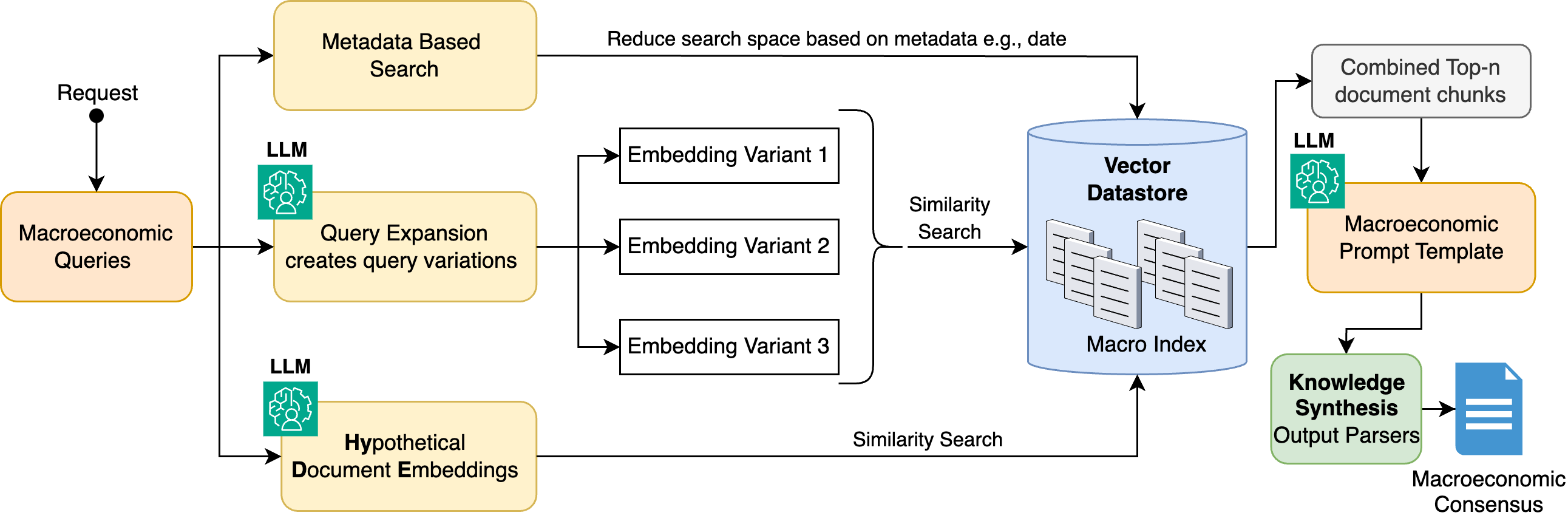}}
    \caption{Macroeconomic Agent's functions during data generation.} 
    \label{fig:macro2}
\end{figure}

As illustrated in Fig.~\ref{fig:macro2}, the \emph{Data Generation} stage transforms user queries into a comprehensive macroeconomic consensus by retrieving, consolidating, and synthesizing relevant information from the vectorized knowledge base. Although MarketSenseAI primarily uses this mechanism to produce concise macro summaries for single-stock analysis, the underlying design also supports broader financial applications, such as powering a conversational assistant or analyzing proprietary research. 

All the input queries to Macroecomic Agent, first undergo metadata filtering, which narrows the set of candidate documents by date or source. From there, retrieval strategies differ based on the use case:

\begin{itemize} 
\item \textbf{MarketSenseAI} (Predefined Queries \& HyDE): For single-stock analysis, we employ a HyDE approach on a fixed set of queries (e.g., “U.S.\ macro outlook,” “investment opportunities and risks”). This yields brief, rounded macroeconomic insights without overburdening the final signal-generation stage. An example of this output is given at Table~\ref{tab:macro-example}.

\item \textbf{Other Use Cases} (Optimized Retrieval with Query Expansion): For open-ended or complex queries, the system uses expanded embeddings and refined prompts to improve coverage, particularly when user requests are ambiguous or partial. By generating multiple query variants, the agent captures broader document matches and delivers more comprehensive responses.
\end{itemize}

\begin{table*}[htbp]
\caption{Macroeconomic Agent's Output (January 3, 2025) used in MarketSenseAI}
\begin{center}
\tiny
\begin{tabular}{|p{2.2cm}|p{15cm}|}
\hline
\textbf{Category} & \textbf{Key Findings} \\
\hline
Global Market Consensus & 
\textbf{US Markets:} S\&P 500 and Nasdaq show significant gains despite Chicago PMI decline and GDP forecast revisions; \textbf{Labor Market:} Remains tight, supporting consumer spending; \textbf{European Markets:} UK benefits from rising housing prices and weaker pound, other major indexes declining; \textbf{Asian Markets:} Japan faces yen weakening, China shows tentative recovery with stimulus; \textbf{Emerging Markets:} Turkey shows resilience, Mexico balancing monetary strategies; \textbf{Bond Market:} Opportunities in high-quality fixed income and green bonds \\
\hline
Contradictory Market Signals &
\textbf{US Economic Indicators:} Strong foreign demand for US securities vs PMI drop and GDP forecast revisions; \textbf{Inflation \& Rates:} Mixed signals affecting monetary policy and market stability, varying expectations for interest rate trajectories; \textbf{Global Performance:} Strong US stocks vs challenges in Japan and China; \textbf{Investment Strategies:} Divergent recommendations between US assets and global diversification; \textbf{Sector Opportunities:} Growth in China's tertiary industry vs US equity resilience \\
\hline
Positive Market Indicators &
\textbf{US Equities:} Strong annual gains in large-cap growth stocks; \textbf{Fixed Income:} Attractive yields and spreads; \textbf{Emerging Markets:} Favorable pricing and long-term potential; \textbf{China Tech:} Strategic emerging industries showing growth potential; \textbf{Sustainability:} Climate innovation offering new investment opportunities \\
\hline
Risk Factors \& Negative Indicators & 
\textbf{US Manufacturing:} Significant drop in Chicago PMI and GDP forecast revisions; \textbf{Japan:} Continued manufacturing contraction and economic uncertainty; \textbf{China:} Declining industrial profits and weak manufacturing data; \textbf{Consumer Metrics:} Decline in US consumer confidence; \textbf{Market Dynamics:} High likelihood of reversal in momentum stocks; \textbf{International Position:} Growing US reliance on foreign capital \\
\hline
\end{tabular}
\label{tab:macro-example}
\end{center}
\end{table*}

After extracting the top-$n$ relevant text chunks via similarity search, we feed them into macroeconomic-focused prompt that guides the LLM to use the information available in the retrieved chunks to response to the input query. This process ensures flexible adaptation to different requirements—from highly targeted stock-specific analyses to more exploratory, institution-wide research queries.

\subsubsection{Retrieval Performance Evaluation}

\label{sec:3.3.3}

To assess the retrieval pipeline’s ability to handle macroeconomic queries of varying complexity, we tested three methods (\emph{Simple}, \emph{Optimized}, and \emph{HyDE}) across different chunk sizes, evaluating \textit{context recall}, \textit{context precision}, \textit{answer relevancy}, and \textit{faithfulness} \cite{ragas_metrics_v0_1_21}. Each approach shapes how queries are transformed before performing semantic similarity searches in the vector database, thereby influencing which top-$n$ chunks are retrieved. The results in Table~\ref{tab:retrieval-performance} demonstrate the effectiveness of different retrieval methods across varying chunk sizes for complex macroeconomic queries. These findings offer several key insights:

\begin{itemize} 
    \item \textbf{Context Precision} remains high ($\ge 0.98$) in all configurations, indicating that even when queries span multiple reports, irrelevant chunks are not in the top-$n$ results. This supports the validity of the data injection presented in Section \ref{sec:3.3.1}.
    \item \textbf{Answer Relevancy} exhibits the greatest variability. Both \emph{HyDE} and \emph{Optimized} augment the query with additional context, improving alignment between the query vector and chunk embeddings. This makes retrieved chunks more likely to address the question which is especially beneficial for broader prompts that require drawing information from multiple sources. 
    \item \textbf{Faithfulness} (i.e., factual accuracy) tends to increase with larger chunk sizes, suggesting that a broader context helps mitigate omissions or misunderstandings. Complex queries, such as identifying contradictory viewpoints across documents, benefit most from expanded chunk sizes. 
    \item \textbf{Simple} retrieval, while occasionally competitive in recall, is consistently weaker in relevancy because it lacks query expansions or concept additions to better match chunks in the vector store. Consequently, it struggles to surface the most pertinent segments for multi-faceted queries.
    \item  \textbf{Increase of Chunks} improves the performance across all methods indicating the high quality of the stored content in the vector database. Retrieval of more chucks seems to be particularly beneficial for questions requiring synthesis of information across multiple reports or identification of subtle differences in economic outlooks.

\end{itemize}

\begin{table}[htbp]
\caption{Performance Comparison of Retrieval Methods by Chunk Size}
\begin{center}
\tiny
\begin{tabular}{|c|c|c|c|c|c|c|}
\hline
\textbf{Top-n} & \textbf{Method} & \textbf{Recall} & \textbf{Precision} & \textbf{Relevancy} & \textbf{Faithfulness} & \textbf{Overall} \\
\hline
\multirow{3}{*}{3} & HyDE & 0.77 & 1.00 & 0.76 & 0.94 & 0.87 \\
\cline{2-7}
& Optimized & 0.67 & 1.00 & 0.75 & 0.89 & 0.83 \\
\cline{2-7}
& Simple & 0.75 & 1.00 & 0.48 & 0.86 & 0.77 \\
\hline
\multirow{3}{*}{5} & HyDE & 0.79 & 0.99 & 0.66 & 0.94 & 0.85 \\
\cline{2-7}
& Optimized & 0.79 & 0.99 & 0.56 & 0.96 & 0.82 \\
\cline{2-7}
& Simple & 0.85 & 0.99 & 0.48 & 0.93 & 0.82 \\
\hline
\multirow{3}{*}{7} & HyDE & 0.91 & 1.00 & 0.66 & 0.98 & 0.89 \\
\cline{2-7}
& Optimized & 0.85 & 1.00 & 0.66 & 0.97 & 0.87 \\
\cline{2-7}
& Simple & 0.86 & 0.99 & 0.57 & 0.95 & 0.84 \\
\hline
\end{tabular}
\label{tab:retrieval-performance}
\end{center}
\end{table}

In practice, the results demonstrate that both HyDE and Optimized methods, especially with more chunks, provide robust frameworks for extracting relevant macroeconomic insights from diverse, large-scale reports. Their superior performance in handling complex queries spanning multiple documents and identifying diverse economic themes makes them particularly well-suited for macroeconomic analysis tasks.

\section{Experiments}
\label{sec:4}
This section details our empirical methodology for evaluating MarketSenseAI's efficacy in stock analysis and rating.

\subsection{Data}
We evaluated MarketSenseAI using stocks from the S\&P 100 and S\&P 500 indices. For S\&P 100 stocks, our analysis covers January 2023 to December 2024, providing a two-year evaluation under varying market conditions. We extended our analysis to the broader S\&P 500 universe for calendar year 2024, when comprehensive data became available for the expanded set of stocks. This approach enables assessment of both the model's long-term consistency through S\&P 100 stocks and its scalability to a larger opportunity set through the S\&P 500 analysis. The input data included:

\begin{itemize}
\item \textbf{Stock-specific data}: Financial news, quarterly statements, SEC filings, earnings call transcripts, and historical price data.
\item \textbf{Macroeconomic Data}: Textual data from investment reports, central bank publications (e.g., Federal Reserve, European Central Bank), and other institutional sources. This included expert analyses, monetary policy discussions, and sector-specific research.
\end{itemize}
Monthly trading signals were generated to align with established portfolio rebalancing practices. The S\&P 500 results for 2024 were analyzed independently to evaluate model generalizability across a broader market universe.

\subsection{Technology Stack}

The GPT-4o model serves as the primary LLM for all processes requiring model inference \cite{hurst2024gpt}, while the system maintains an LLM-agnostic architecture that allows seamless integration of alternative models via API. For portfolio analysis and strategy validation, we utilized VectorBTPro, which provided robust tools for backtesting financial strategies while accounting from transaction costs. To assess the RAG methods outlined in Section \ref{sec:3.3}, we employed the Ragas framework \cite{es2023ragas}, leveraging GPT-4o-mini for cost efficiency. While this choice may have impacted evaluation results compared to the full-scale GPT-4o model, this did not affect the relative comparison of the methods under evaluation.

The vector datastore is based on Pinecone and the agents within the system are built on OpenAI’s client. The RAG processes leverage the LlamaIndex framework.

For data collection, macroeconomic reports are scraped using tools like Playwrite, combined with custom scripts tailored to specific data sources. SEC filings are sourced directly from the SEC’s EDGAR API, while earnings call transcripts are obtained via RapidAPI, which aggregates data from platforms such as SeekingAlpha and MarketBeat.

\subsection{Evaluation Approach}

In addition to the agent-specific evaluation presented in Section~\ref{sec:3}, we evaluate the quality of MarketSenseAI's signals by constructing portfolios and comparing their performance against relevant benchmarks. Specifically, we focus on long-only portfolios based on MarketSenseAI's \textit{buy} signals, implemented in two forms: equally weighted and market capitalization weighted. These portfolios are compared against their corresponding equally or market cap weighted benchmark (S\&P 100 or S\&P 500) to assess the system's effectiveness in generating actionable investment signals. The evaluated signals/strategies and their relevant benchmarks are presented in Table~\ref{tab:portfolios}. Typical performance and risk metrics were used for assessing both the MarketSenseAI-based and the benchmark portfolios, including: total return (cumulative portfolio returns over the period), Sharpe ratio (risk-adjusted return relative to volatility), Sortino ratio (return relative to downside risk), volatility (standard deviation of returns), win rate (percentage of profitable trades), and maximum drawdown (MDD, peak-to-trough portfolio loss).

\begin{table}[htbp]
\caption{Investment Strategies and Benchmark Portfolios}
\begin{center}
\tiny
\begin{tabular}{|p{1.5cm}|p{5.5cm}|}
\hline
\textbf{Abbreviation} & \textbf{Description} \\
\hline
MS & Equally weighted portfolio rebalanced monthly based on the \textit{buy} signals of MarketSenseAI \\
\hline
MS-Cap & Capitalization-weighted portfolio rebalanced monthly based on the \textit{buy} signals of MarketSenseAI \\
\hline
S\&P100-Eq & Equally weighted portfolio of all the stocks of the S\&P 100 index (tracked by the EQWL ETF) \\
\hline
S\&P100 & Capitalization-weighted S\&P 100 index (tracked by the OEF ETF) \\
\hline
S\&P500-Eq & Equally weighted portfolio of all the stocks of the S\&P 500 index (tracked by the RSP ETF) \\
\hline
S\&P500 & Capitalization-weighted S\&P 500 index (tracked by the SPY ETF) \\
\hline
\end{tabular}
\label{tab:portfolios}
\end{center}
\end{table}

\section{Results}
\label{sec:5}
This section evaluates MarketSenseAI's stock selection capability through empirical testing on the S\&P 100 (2023-2024) and S\&P 500 (2024) universes. Results demonstrate the system's ability to identify outperforming equities, generating superior risk-adjusted returns across different portfolio construction methodologies.

\subsection{Overall Performance Overview}

MarketSenseAI’s ability to identify outperforming equities is evident across multiple dimensions. In the S\&P 100 universe, the system’s selected stocks achieved a 125.9\% cumulative return under market cap-weighting (MS-Cap), significantly surpassing the S\&P 100 index return of 73.5\% (Table~\ref{tab:strategy_metrics}). This outperformance persisted in equal-weighted portfolios (MS-Eq), where selected equities returned 55.7\% versus 42.3\% for the equal-weighted S\&P 100. Critically, these gains were not achieved through excessive risk-taking: the MS-Cap portfolio exhibited a 16\% higher Sortino ratio (4.43 vs. 3.82) compared to the cap-weighted benchmark, despite experiencing higher volatility.

\begin{table}[htbp]
\caption{Performance Metrics (2023-2024)}
\begin{center}
\tiny
\begin{tabular}{|p{0.9cm}|p{0.8cm}|p{0.5cm}|p{0.6cm}|p{0.5cm}|p{0.5cm}|p{0.6cm}|}
\hline
\textbf{Portfolio} & \textbf{Return$^{\mathrm{a}}$} & \textbf{Sharpe} & \textbf{Sortino} & \textbf{Vol} & \textbf{MDD} & \textbf{MDDd$^{\mathrm{b}}$} \\
\hline
\multicolumn{7}{l}{\textbf{S\&P 100 Analysis (2023-2024)}} \\
\hline
MS-Eq & 55.7 (53.2) & 2.13 & 3.25 & 15.6 & 9.2 & 65 \\
\hline
S\&P 100 Eq & 42.3 (42.3) & 1.89 & 2.85 & 14.1 & 10.7 & 92 \\
\hline
MS-Cap & 125.9 (123.0) & 2.76 & 4.43 & 22.3 & 13.8 & 82 \\
\hline
S\&P 100 & 73.5 (73.5) & 2.52 & 3.82 & 16.4 & 9.7 & 77 \\
\hline
\multicolumn{7}{l}{\textbf{S\&P 500 Analysis (2024)}} \\
\hline
MS-Eq & 25.8 (24.5) & 2.4 & 3.68 & 14.3 & 6.7 & 52 \\
\hline
S\&P 500 Eq & 12.8 (12.8) & 1.33 & 1.91 & 13.8 & 7.1 & 73 \\
\hline
MS-Cap & 48.7 (47.8) & 2.87 & 4.39 & 20.8 & 12.5 & 53 \\
\hline
S\&P 500 & 25.6 (25.6) & 2.26 & 3.28 & 15.1 & 8.4 & 46 \\
\hline
\multicolumn{7}{p{7cm}}{$^{\mathrm{a}}$Values in parentheses represent the total returns (\%) after transaction costs (10bps/trade).} \\
\multicolumn{7}{p{7cm}}{$^{\mathrm{b}}$The duration of Maximum Drawdown (MDD) in days.}
\end{tabular}
\label{tab:strategy_metrics}
\end{center}
\end{table}

The system’s selection capability scaled effectively with market breadth. When applied to the S\&P 500 universe during 2024, MarketSenseAI’s selected equities delivered 25.8\% returns in equal-weighted portfolios compared to 12.8\% for the S\&P 500 Equal Weight benchmark, representing a 102\% relative outperformance. This expansion to a broader universe also improved risk-adjusted performance, with the Sortino ratio increasing from 3.25 (S\&P 100 MS-Eq) to 3.68 (S\&P 500 MS-Eq). Alpha generation improved correspondingly, rising from 8.0\% in the S\&P 100 MS-Eq to 18.9\% in the S\&P 500 MS-Eq (Table~\ref{tab:attribution}), confirming the system’s enhanced ability to identify opportunities in larger universes.

\begin{table}[htbp]
\caption{Performance Attribution Analysis}
\begin{center}
\tiny
\begin{tabular}{|p{1.0cm}|p{0.6cm}|p{0.8cm}|p{0.7cm}|p{0.8cm}|p{1.4cm}|}
\hline
\textbf{Portfolio} & \textbf{Beta} & \textbf{Alpha (\%)} & \textbf{Total Trades} & \textbf{Win Rate (\%)} & \textbf{Buy Signals$^{\mathrm{a}}$} \\
\hline
\multicolumn{6}{l}{\textbf{S\&P 100 Analysis (2023-2024)}} \\
\hline
MS-Eq & 0.96 & 8.0 & 584 & 77.1 & 35.1 (7.95) \\
\hline
MS-Cap & 1.24 & 10.6 & 548 & 77.0 & 35.1 (7.95) \\
\hline
\multicolumn{6}{l}{\textbf{S\&P 500 Analysis (2024)}} \\
\hline
MS-Eq & 0.92 & 18.9 & 1200 & 78.0 & 144.8 (30.8) \\
\hline
MS-Cap & 1.27 & 17.6 & 1229 & 77.0 & 144.8 (30.8) \\
\hline
\multicolumn{6}{p{7cm}}{$^{\mathrm{a}}$Values in the parentheses represent the standard deviation of the average number of buy signals per month.}
\end{tabular}
\label{tab:attribution}
\end{center}
\end{table}

\begin{figure}[!t]
\centering
\begin{subfigure}{0.48\columnwidth}
\centering
\includegraphics[width=\textwidth]{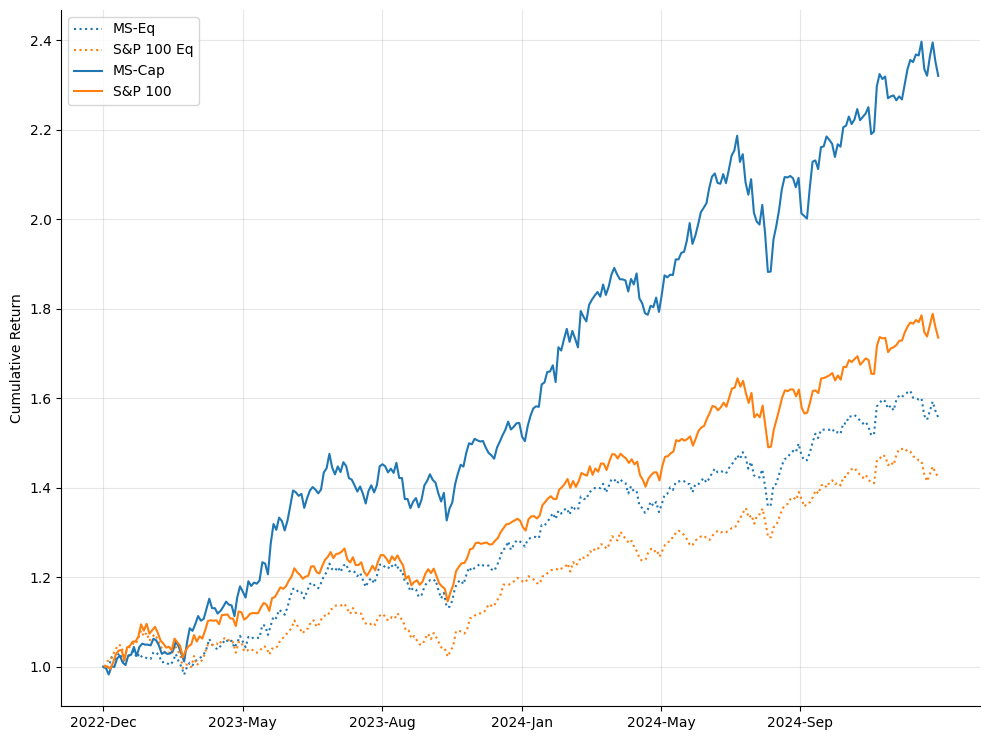}\label{fig:spx100}
\caption{}
\end{subfigure}
\hfil
\begin{subfigure}{0.48\columnwidth}
\centering
\includegraphics[width=\textwidth]{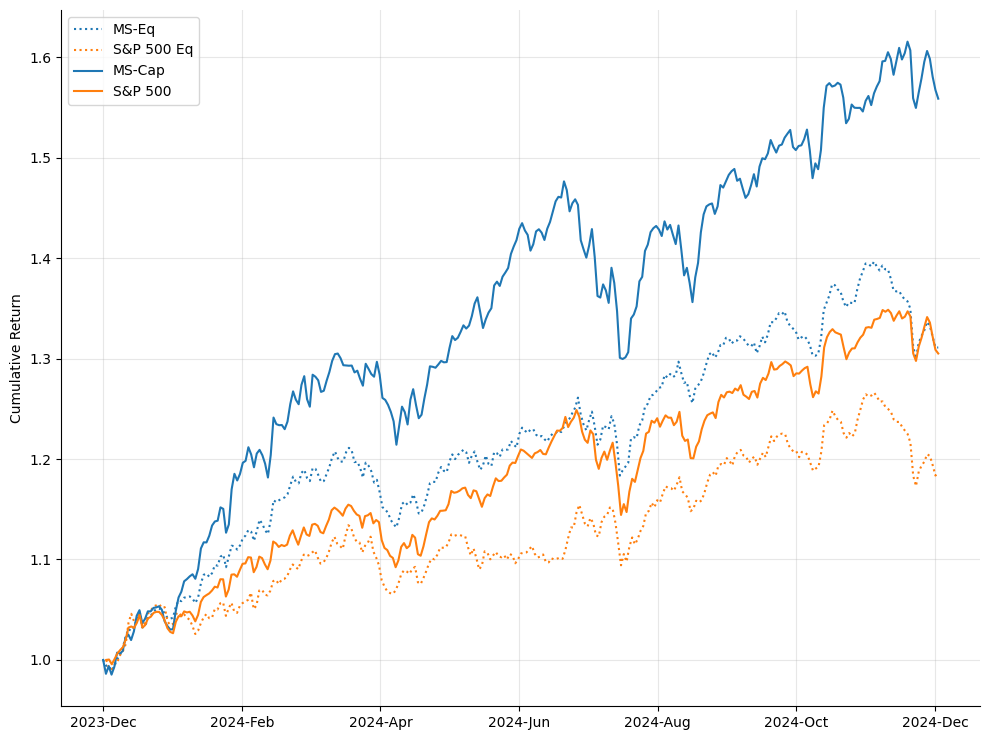}\label{fig:spx500}
\caption{}
\end{subfigure}
\caption{Cumulative returns of MarketSenseAI \textit{buy} monthly signals against the market: (a) S\&P 100 stock universe (2023-2024) and (b) S\&P 500 stock universe (2024).
\label{fig:perfomance}}
\end{figure}

Furthermore, despite selecting higher-volatility equities, MarketSenseAI-based portfolios recovered quite fast from drawdowns, while maintaining a comparable maximum drawdowns with the benchmarks. This resilience is visually confirmed in Fig.~\ref{fig:perfomance}, where the system’s cumulative returns exhibit fast recoveries during market stress periods yet with an upward trend.

The attribution analysis (Table~\ref{tab:attribution}) reveals additional insights. With a 77–78\% win rate across implementations, the system demonstrates remarkable consistency in signal precision. The positive alpha (17.6–18.9\%) and elevated beta (1.24–1.27) of the S\&P 500 portfolios suggest MarketSenseAI successfully identifies high-beta stocks with idiosyncratic upside potential. Furthermore, the stable monthly signal generation—35.1 ±7.95 buy signals for S\&P 100 and 144.8 ±30.8 for S\&P 500—indicates systematic selection rather than concentrated bets.


\subsection{Factor Analysis and Risk Decomposition}
To elucidate the drivers of MarketSenseAI’s outperformance, we decompose portfolio returns (MS-Eq - S\&P 100) using the Carhart four-factor \cite{Carhart1997} and Fama-French five-factor \cite{FamaFrench2015} models. Both models explain a substantial portion of return variance ($R^2 = 88.4\%$ and $85.4\%$, respectively), validating their applicability. Key findings are summarized in Table \ref{tab:factor_results} and discussed below.

\subsubsection{Market Exposure and Size Bias}
MarketSenseAI exhibits near-neutral market exposure ($\beta = 0.95$--$0.96$). The negative SMB coefficients ($-0.13$ to $-0.22$, $p < 0.01$) reflect a tilt toward large-cap stocks, aligning with the S\&P 100/500 universes\footnote{SMB: size premium (small minus big capitalization stocks).}.

\subsubsection{Value and Momentum Factors}
Both models confirm consistent value exposure (HML = 0.08--0.11, $p < 0.01$)\footnote{HML: value premium (high minus low book-to-market ratio stocks).}, underscoring the Fundamentals Agent’s ability to identify undervalued equities through financial statement analysis. The Carhart model’s strong momentum loading (Mom = 0.18, p < 0.01) highlights MarketSenseAI’s integration of price trends via the Dynamics Agent, a feature often absent in traditional fundamental models. This synergy between value and momentum aligns with the system’s architecture, where LLM-driven news sentiment and price dynamics reinforce fundamental insights.

\subsubsection{Profitability and Investment Factors}
The five-factor model reveals insignificant loadings on profitability (RMW) and investment (CMA)\footnote{RMW: profitability premium (robust minus weak firms); CMA: investment premium (conservative minus aggressive investment policies).}, suggesting these factors play minimal roles in MarketSenseAI's strategy. This suggests MarketSenseAI's returns are not systematically driven by these traditional style factors. The system's integration of multiple data sources may help identify alpha sources beyond conventional factor premiums.

\subsubsection{Alpha Generation and Unexplained Returns} 
The analysis reveals a significant residual alpha ($+8.0\%$, Table \ref{tab:attribution}) and substantial unexplained returns ($12$--$15\%$) that cannot be attributed to traditional risk factors. These results suggest potential value generation beyond conventional factor exposure, they may reflect MarketSenseAI's consideration of multiple data sources such as news narratives, macroeconomic context, and forward-looking disclosures which enable the identification of idiosyncratic opportunities overlooked by factor-based models.

\begin{table}[htbp]
\caption{Factor Model Results}
\begin{center}
\tiny
\begin{tabular}{|l|c|c|}
\hline
\textbf{Factor} & \textbf{Carhart 4-Factor} & \textbf{Fama-French 5-Factor} \\
\hline
$\text{Mkt-RF}$ ($\beta$) & 0.936$^{***}$ & 0.958$^{***}$ \\
\hline
$\text{SMB}$ & -0.131$^{***}$ & -0.221$^{***}$ \\
\hline
$\text{HML}$ & 0.110$^{***}$ & 0.081$^{***}$ \\
\hline
$\text{Mom}$ & 0.178$^{***}$ & -- \\
\hline
$\text{RMW}$ & -- & -0.015 \\
\hline
$\text{CMA}$ & -- & 0.044 \\
\hline
$R^2$ & 0.884 & 0.854 \\
\hline
\multicolumn{3}{l}{Note: $^{***}p < 0.01$, $^{**}p < 0.05$, $^{*}p < 0.1$. Dashes (--)} indicate\\
\multicolumn{3}{l}{factor not included in model. Market factor (Mkt-RF) shows} near-\\
\multicolumn{3}{l}{unity exposure, SMB reflects large-cap bias, HML} demonstrates\\
\multicolumn{3}{l}{value exposure, and Mom captures momentum effects.}
\end{tabular}
\label{tab:factor_results}
\end{center}
\end{table}



\section{Conclusions}
\label{sec:6}

This paper presented significant advancements in the MarketSenseAI framework, demonstrating the efficacy of integrating LLM agents and retrieval-augmented techniques for holistic stock analysis. By addressing critical challenges such as context window limitations, data frequency mismatches, and the integration of qualitative and quantitative information, the framework introduced  a Chain-of-Agents approach for granular fundamental analysis and a RAG module enhanced with HyDE for macroeconomic context. These advancements enable deeper, more comprehensive analysis of SEC filings, earnings calls, and expert reports, which traditional models often overlook.

Empirical evaluations on S\&P 100 (2023–2024) and S\&P 500 (2024) stocks validate MarketSenseAI’s efficacy. While the S\&P 500 analysis was limited to 2024 due to data availability, the system’s ability to scale to a larger universe (500 stocks) while improving performance underscores its robust stock-picking capabilities. The framework generated significantly higher cumulative returns and consistent alpha, outperforming competitive benchmarks across risk-adjusted metrics. Factor analysis revealed that returns stem not only from exposure to value and momentum factors but also from unique alpha sources, likely attributable to the framework’s versatile data integration and analysis.

Future development will focus on two key directions: technological advancement through integration of reasoning-enabled LLMs and market expansion to global and small-cap indices. These enhancements aim to further improve the system's analytical capabilities while testing its adaptability across diverse market conditions.

MarketSenseAI represents a significant step forward in applying LLMs to financial analysis, offering both institutional and retail investors a transparent, data-driven approach to investment decision-making. By successfully addressing fundamental challenges such as processing lengthy documents, mitigating hallucination risks, and integrating multiple data sources, this work establishes a foundation for building more intelligent investment frameworks.

\bibliography{bibliography}
\bibliographystyle{IEEEtran}

\end{document}